# Cuprate-like Electronic Structures in Infinite-Layer Nickelates with Substantial Hole Dopings

Xiang Ding[1]†, Yu Fan[1]†, Xiaoxiao Wang[1], Chihao Li[1], Zhitong An[1], Jiahao Ye[1], Shenglin Tang[1], Minyinan Lei[1], Xingtian Sun[1], Nan Guo[1], Zhihui Chen[1], Suppanut Sangphet[1], Yilin Wang[2,3], Haichao Xu[1,4]*, Rui Peng[1,4]*, Donglai Feng[5,1,2,3]*

[1] Advanced Materials Laboratory, State Key Laboratory of Surface Physics, and Department of Physics, Fudan University, Shanghai 200433, China

[2] School of Emerging Technology, University of Science and Technology of China, Hefei 230026, China

[3] New Cornerstone Science Laboratory, University of Science and Technology of China, Hefei 230026, China

[4] Shanghai Research Center for Quantum Sciences, Shanghai 201315, China.

[5] National Synchrotron Radiation Laboratory and School of Nuclear Science and Technology, University of Science and Technology of China, Hefei 230026, China

*Corresponding authors. Emails: xuhaichao@fudan.edu.cn, pengrui@fudan.edu.cn, dlfeng@ustc.edu.cn.

† These authors contributed equally to this work.

## ABSTRACT

The superconducting infinite-layer (IL) nickelates offer a new platform for investigating the long-standing problem of high-temperature superconductivity. Many models were proposed to understand its superconducting mechanisms based on the calculated electronic structure, and the multiple Fermi surfaces and multiple orbitals involved create complications and controversial conclusions. Over the past 5 years, the lack of direct measurements of the electronic structure has hindered the understanding of nickelate superconductors. Here we fill this gap by directly resolving the electronic structures of the parent compound $LaNiO_2$ and superconducting $La_{0.8}Ca_{0.2}NiO_2$ using angle-resolved photoemission spectroscopy (ARPES). We find that their Fermi surfaces consist of a quasi-two-dimensional (quasi-2D) hole pocket and a three-dimensional (3D) electron pocket at the Brillouin zone corner, whose volumes change upon Ca doping. The Fermi surface topology and band dispersion of the hole pocket closely resemble those observed in hole-doped cuprates. However, the cuprate-like band exhibits significantly higher hole doping in superconducting $La_{0.8}Ca_{0.2}NiO_2$ compared to superconducting cuprates, highlighting the disparities in the electronic states of the superconducting phase. Our observations highlight the novel aspects of the IL nickelates, and pave the way toward the microscopic understanding of the IL nickelate family and its superconductivity.

## INTRODUCTION

Following the discovery of the high-temperature superconductivity in cuprates[1], it was suggested that superconductivity in nickelates could be realized, if the common Ni $3d^8$ state could be reduced to $3d^9$ [2-6]. Indeed, superconductivity was discovered in (Nd,Sr)$NiO_2$ thin films[7], after the apical oxygens in (Nd,Sr)$NiO_3$ films were removed by reaction with $CaH_2$ powders. Subsequently, superconductivity was achieved in related compounds such as (La,Sr)$NiO_2$[8,9], (La,Ca)$NiO_2$[10], (Pr,Sr)$NiO_2$[11], (Nd,Eu)$NiO_2$[12], etc. However, high-quality IL nickelate superconductors are difficult to fabricate[13-15], and the surface of the IL nickelate superconductors usually becomes disordered in the reduction process[16,17], which prevents the reliable measurement of its electronic structure by ARPES or scanning tunneling microscopy (STM).

Various theoretical models on the superconductivity of IL nickelates are based on combinations of different Ni/*RE* (rare earth) orbitals and Fermi surface topologies. Consequently, distinct superconducting mechanisms could be reached. For instance, Kitatini et.al. propose that *RE*$NiO_2$ can be described by one band Hubbard model with Ni-3$d_{x^2-y^2}$ orbital akin to cuprates, based on which superconducting transition temperature can be estimated[18]. However, others suggested that Ni-3$d_{xy}$ or Ni-3$d_{3z^2-r^2}$ orbital and Hund's coupling should be included, potentially yielding a high-spin $S = 1$ state in superconducting nickelates[19,20]. Additionally, the presence of conduction electrons (including various *RE-d* orbitals and interstitial *s* orbitals) and their contributions to superconductivity further complicate the understanding[21-28]. Depending on different hybridization and doping levels, various pairing symmetries distinct from hole-doped cuprates were predicted [29-31].

Since accurate knowledge of the low-energy electronic structure is critical for modeling the IL nickelates, many fundamental issues need to be pinned down, such as the Fermi surface topology, hole concentration, the orbital characters of bands, the participation of *RE-5d* or interstitial *s* orbitals in the low energy electronic structure, etc. Particularly, a key question is whether the electronic structure resembles those of cuprates. However, due to strong electron correlations, an accurate band calculation for IL nickelates is still challenging, thus direct experimental studies are demanded.

# RESULTS

## Single-crystalline IL surface

Reliable measurements of the electronic structure of IL nickelates require high-quality stoichiometric $RENiO_3$ perovskite films, sufficient *in-situ* reduction, and most critically, single crystalline IL surfaces. Especially, achieving single crystalline IL surfaces poses a significant challenge, as it requires a delicate balance in the strength of the reduction conditions: strong enough to facilitate a topotactic transition to the IL phase, yet mild enough to prevent damage to the crystalline surface, which is essential for surface-sensitive techniques like ARPES. To address this challenge, we have performed *in-situ* reduction and systematically optimized the reducing conditions. We have grown $LaNiO_3$ and $La_{0.8}Ca_{0.2}NiO_3$ thin films on $SrTiO_3$ (001) substrates using oxide molecular beam epitaxy with an atomic-layer-by-layer growth method (See Materials and Methods in Supplementary Materials), and then reduced them *in-situ* with atomic hydrogen. As depicted in Fig. 1A, we have used a shutter to prevent direct H atom bombardment on the sample surface, which effectively avoids disorder formation during the violent topotactic reduction process. In this way, the reflection high energy electron diffraction (RHEED) pattern of the reduced films shows sharp streaks from the surfaces (Fig. 1A), and atomic force microscopy (AFM) shows terraces with unit-cell step height (Supplementary Fig. S4), indicating single-crystalline and atomically flat sample surfaces. *Ex-situ* X-ray diffraction (XRD) measurements were performed on the same samples after ARPES measurements (Fig. 1B and Fig. S3). The positions of the diffraction peaks shift to higher values compared to those of the perovskite phase (Fig. 1B), and are consistent with those of $(La,Ca)NiO_2$[10]. The fringes accompanying the diffraction peaks observed in $La_{0.8}Ca_{0.2}NiO_2$ (Fig. 1B) and $LaNiO_2$ (Supplementary Fig. S3A) are comparable to, if not more pronounced than, the previous reports[8-10,17]. The conversion efficiency from perovskite to IL phase is among the highest as compared to literature[17](Supplementary Text). These results demonstrate the acquisition of the IL phase with superior surface quality. According to the resistivity measurements (Fig.1C), 21 unit cell (uc) $LaNiO_2/SrTiO_3$ shows a weakly-insulating behavior below 25 K, while 25 uc $La_{0.8}Ca_{0.2}NiO_2/SrTiO_3$ shows a superconducting transition at 8 K. These behaviors are in line with the reported phase diagram of $(La,Ca)NiO_2$ (Fig. 1D, ref. [10]). ARPES measurements on these samples show clear Fermi surfaces (Figs. 1E-1F).

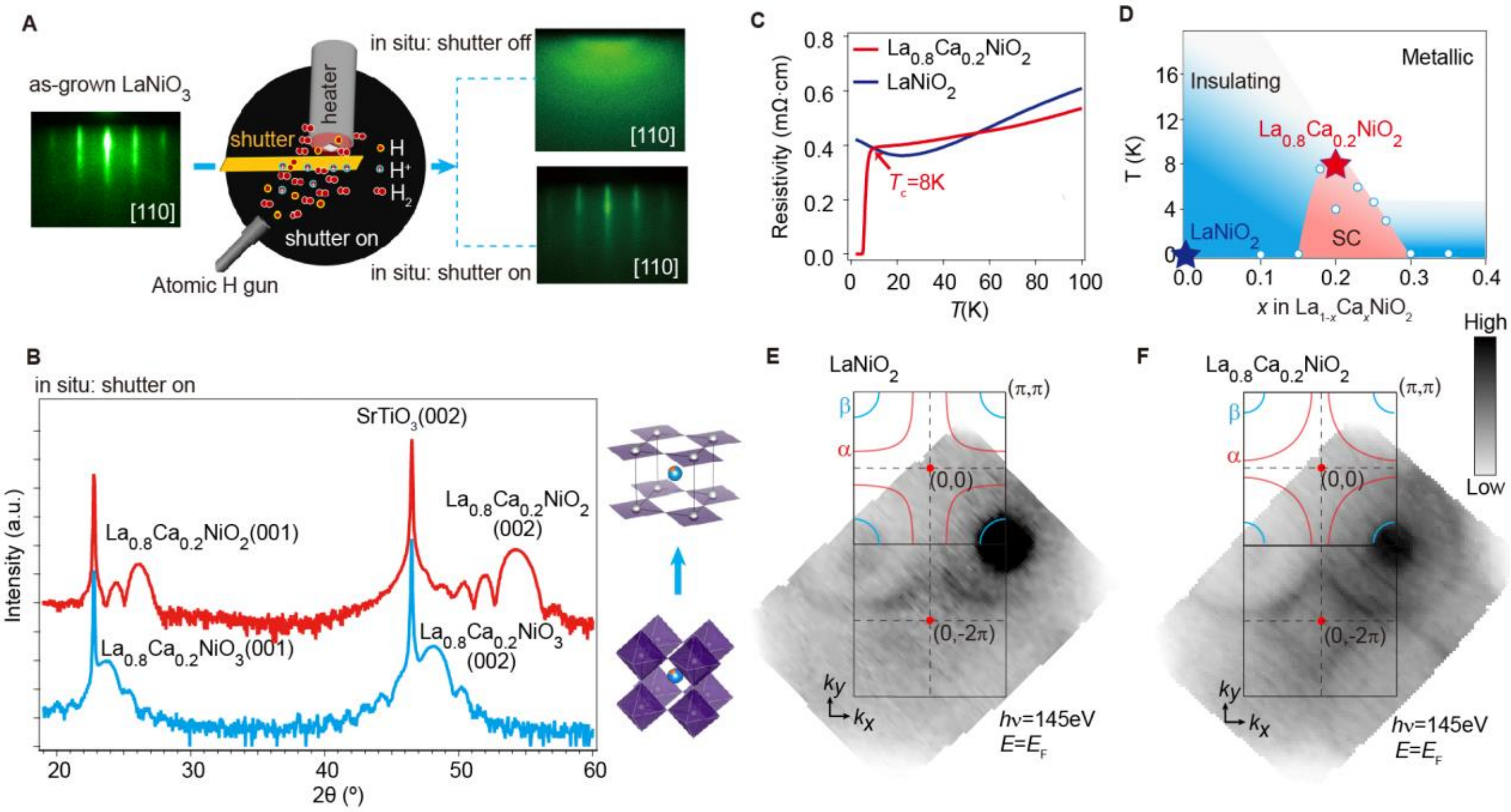

**Fig. 1. *In-situ* reduction and optimization to get atomically flat and clean surfaces for ARPES measurements.** (**A**) Evolution of RHEED image along the [110] azimuth after *in- situ* reduction. The RHEED pattern is single-crystalline (poly-crystalline) when the shutter is on (off) after reduction. The shutter was designed to screen the by-product $H^+$ generated with atomic H. (**B**) XRD $\theta$-$2\theta$ scans of the perovskite 25 uc $La_{0.8}Ca_{0.2}NiO_3/SrTiO_3$ and *in-situ* reduced IL $La_{0.8}Ca_{0.2}NiO_2/SrTiO_3$. (**C**) Temperature-dependent resistivity curves of the $LaNiO_2/SrTiO_3$ and $La_{0.8}Ca_{0.2}NiO_2/SrTiO_3$ samples in this study. (**D**) Superconducting $T_c$ vs. Ca doping level plot in the phase diagram of (La,Ca)$NiO_2$ adapted from Ref.10. The open circles represent data points reported in Ref. [10], while the filled stars illustrate the data obtained from our samples. (**E, F**) Photoemission intensity map of 21 uc $LaNiO_2/SrTiO_3$ and 25 uc $La_{0.8}Ca_{0.2}NiO_2/SrTiO_3$ at $E_F$. The integration is over the energy window of $E_F \pm 0.1$ eV. The red rounded rectangular and the blue small pockets are denoted as $\alpha$ and $\beta$ pockets, respectively.

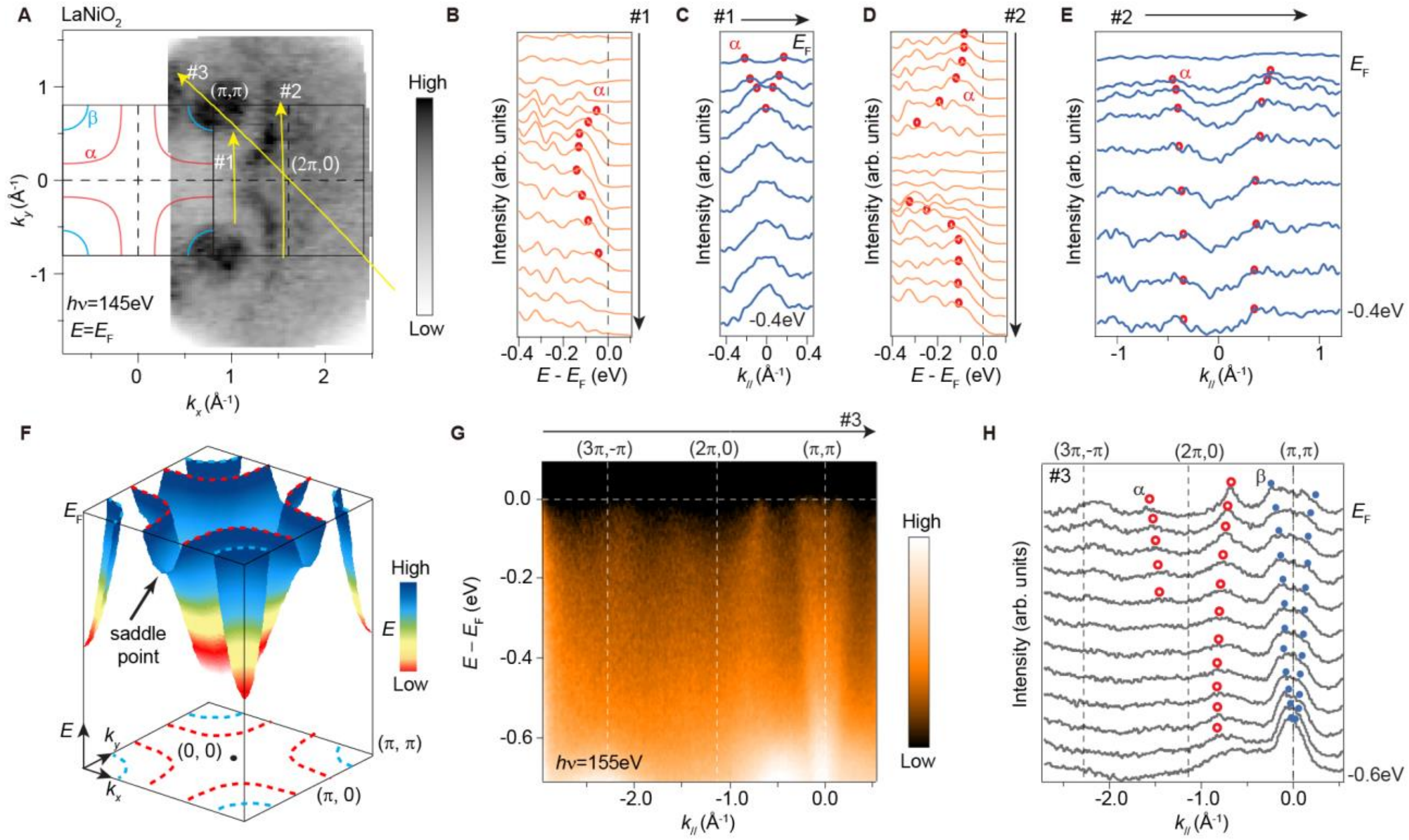

**Fig. 2. Cuprate-like low-energy electronic structure of $LaNiO_2$.** (**A**) Photoemission intensity map of 21uc $LaNiO_2/SrTiO_3$ at $E_F$. The integration is over the energy window of $E_F \pm 0.1$ eV. The Fermi surfaces of a cuprate-like hole pocket $\alpha$ and an electron pocket $\beta$ are illustrated. (**B**) Energy distribution curves (EDCs) along the momentum cut #1. (**C**) Momentum distribution curves (MDCs) along the momentum cut #1. (**D**) the same as panel B, but along cut #2. (**E**) the same as panel C, but along cut #2. (**F**) Schematic dispersion of $\alpha$ and $\beta$ bands. The saddle point of the α band is indicated. (**G**) Photoemission intensity along cut #3. (**H**) MDCs along cut #3. The circle markers track the local maxima/shoulders to demonstrate the dispersion of $\alpha$ band (red circles) and $\beta$ band (blue circles).

## Cuprate-like band dispersion

Figure 2 shows the detailed electronic structure of $LaNiO_2$ measured by ARPES. The $\alpha$ band resembles the low-energy Zhang-Rice singlet of cuprates in terms of both Fermi surface shape and band dispersion [32]. It forms a large rounded square pocket centered at ($\pi$, $\pi$) with parallel sectors around ($\pi$, 0) (Fig. 2A). Note that the spectral weight intensity is higher in the second Brillouin zone (BZ) (Figs.1E,1F,2A), a phenomenon commonly observed in ARPES studies of cuprates and consistent with the photoemission matrix-element of $3d_{x^2-y^2}$ orbitals, and our polarization dependent ARPES measurements also support its $d_{x^2-y^2}$ character (as shown in the Fig.S5 of Supplementary Materials). The dispersion of $\alpha$ band exhibits a shallow electron-like dispersion along cut #1 in Figs. 2B and 2C. Along cut #2, the dispersion of the $\alpha$ band is steep near the zone center (Fig. 2E) and it flattens towards lower binding energy near (0, $\pi$) (Fig. 2D). These demonstrate a saddle-point dispersion in the (0, $\pi$) region (Fig. 2F), akin to the dispersion in the anti-nodal region of cuprates [32]. The saddle-point dispersion is also observed in $La_{0.8}Ca_{0.2}NiO_2$ (Supplementary Fig.S6), highlighting its similarity to cuprates. On the other hand, there is an electron-like band centered at ($\pi$, $\pi$) (Figs. 2G-2H), which is absent in cuprates. Note that the quasiparticle weight is weak but discernible here, manifested by the sharper peaks in the momentum distribution curves near Fermi level (Fig.2H) and reduced band velocity as the energy approaches the Fermi energy (Fig.2G). The spectral width of the α band broadens as the energy moves away from the Fermi level, and the dispersion of the $\alpha$ band becomes steeper at binding energies beyond 0.2 eV, similar to the steep "water-fall" dispersion observed in the cuprates[33].

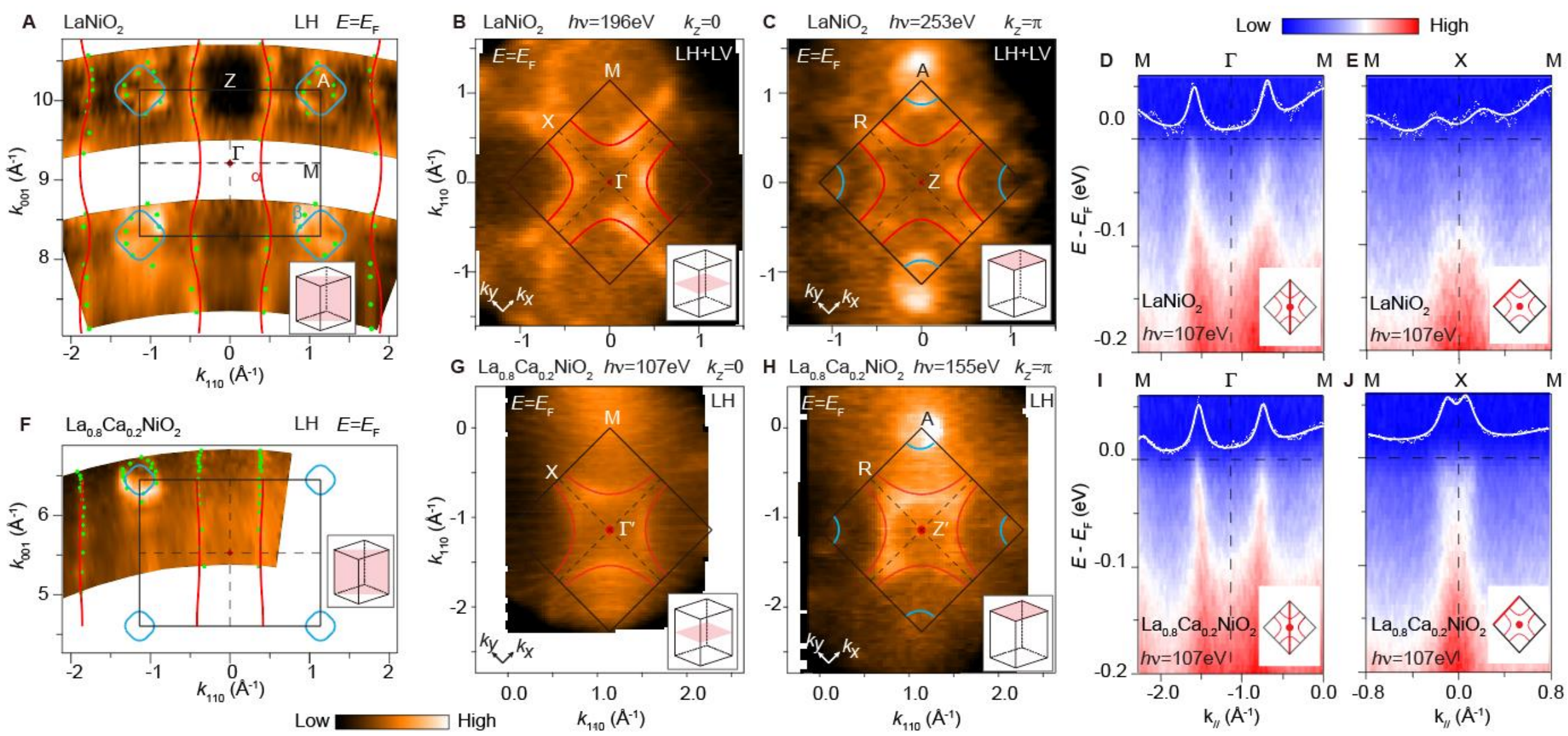

**Fig. 3. Doping dependence of the three-dimensional electronic structure.** (**A**) Photon-energy dependent photoemission intensity map of $LaNiO_2/SrTiO_3$ at $E_F$ in the Γ-M-A-Z plane. (**B**) Photoemission intensity map at $E_F$ measured at Γ-M-X plane ($k_z$=0). (**C**) Same as panel B but measured at the Z-A-R plane ($k_z$=$\pi$). The linear horizontal (LH) and linear vertical (LV) polarizations of photons are indicated, corresponding to the photoemission geometry of $\pi$-polarization and σ-polarization, respectively. (**D-E**) Photoemission spectra along the M-Γ-M direction (D) and M-X-M direction (E) of $LaNiO_2/SrTiO_3$ measured using 107 eV photons ($k_z$=0). The MDCs at $E_F$ were overlaid to show the Fermi crossings. (**F-J**) Same as panels A-E but measured on $La_{0.8}Ca_{0.2}NiO_2/SrTiO_3$.

### Distinct hole doping phase diagram

To reveal the three-dimensional electronic structure of IL nickelates, we further conducted photon energy-dependent ARPES measurements on the films. As shown in Figs. 3A and 3F, the $\alpha$ band shows weak dispersion along $k_z$, demonstrating its quasi-two-dimensional character, whereas the $\beta$ band is three-dimensional and only appears at the A point. In the Γ-M-X plane, the $\alpha$ Fermi surface forms a large hole pocket centered at M (Figs. 3B, 3G), consistent with theoretical calculations [34-36] (Fig.S8). The electron pocket with dominant La-5$d_{3z^2-r^2}$ character predicted at Γ is absent in the experiment (Figs.3B, 3G). In Figs.3C and 3H, the circular pocket formed by the $\beta$ band is identified around the A point. It's noteworthy that the Fermi surface of the $\alpha$ band in the Z-A-R plane roughly matches that in the Γ-M-X plane, consistent with its quasi-two-dimensional character. This is different from the prediction by DFT calculations, where the hole pockets of the $\alpha$ band around A expand and transform into an electron pocket around Z in the Z-A-R plane [34].

As a function of the hole doping in the Zhang-Rice singlet band, cuprate superconductors show a generic phase diagram across various families of materials [37]. Here we compare the doping of the cuprate-like $\alpha$ band of Ni-3$d_{x2-y2}$ character in IL nickelates with the general phase diagram of cuprates. According to the Luttinger theorem and the measured Fermi surface volume, the quasi-two-dimensional Fermi surface of the $d_{x2-y2}$ band possesses 1.09 holes in $LaNiO_2$ (see Supplementary Text for details), indicating an excess of 0.09 holes relative to the $3d^9$ electronic configuration, far from the half-filled Mott insulator. Upon Ca substitution, the $\beta$ pocket also shrinks (Fig. S7), and the Fermi crossings of the $\alpha$ band shift away from the M point along both the M-Γ-M direction (Figs.3D,3I) and M-X-M direction (Figs.3E,3J), indicating an increase of the hole pocket size. The estimated hole concentration is 1.28 for the $\alpha$ pocket of $La_{0.8}Ca_{0.2}NiO_2$. Therefore, despite the resemblance in the band dispersion, the cuprate-like $\alpha$ band in optimally doped $La_{0.8}Ca_{0.2}NiO_2$ possesses an ultra-high doping level of 28%, placing it in the over-doped and non-superconducting regime of the cuprates [37]. As illustrated in the phase diagram (Fig.4A), the superconducting dome of IL nickelates shows up at a higher doping regime of the $d_{x2-y2}$ band than that of cuprates, which highlights the intriguing differences between the nickelate and cuprates.

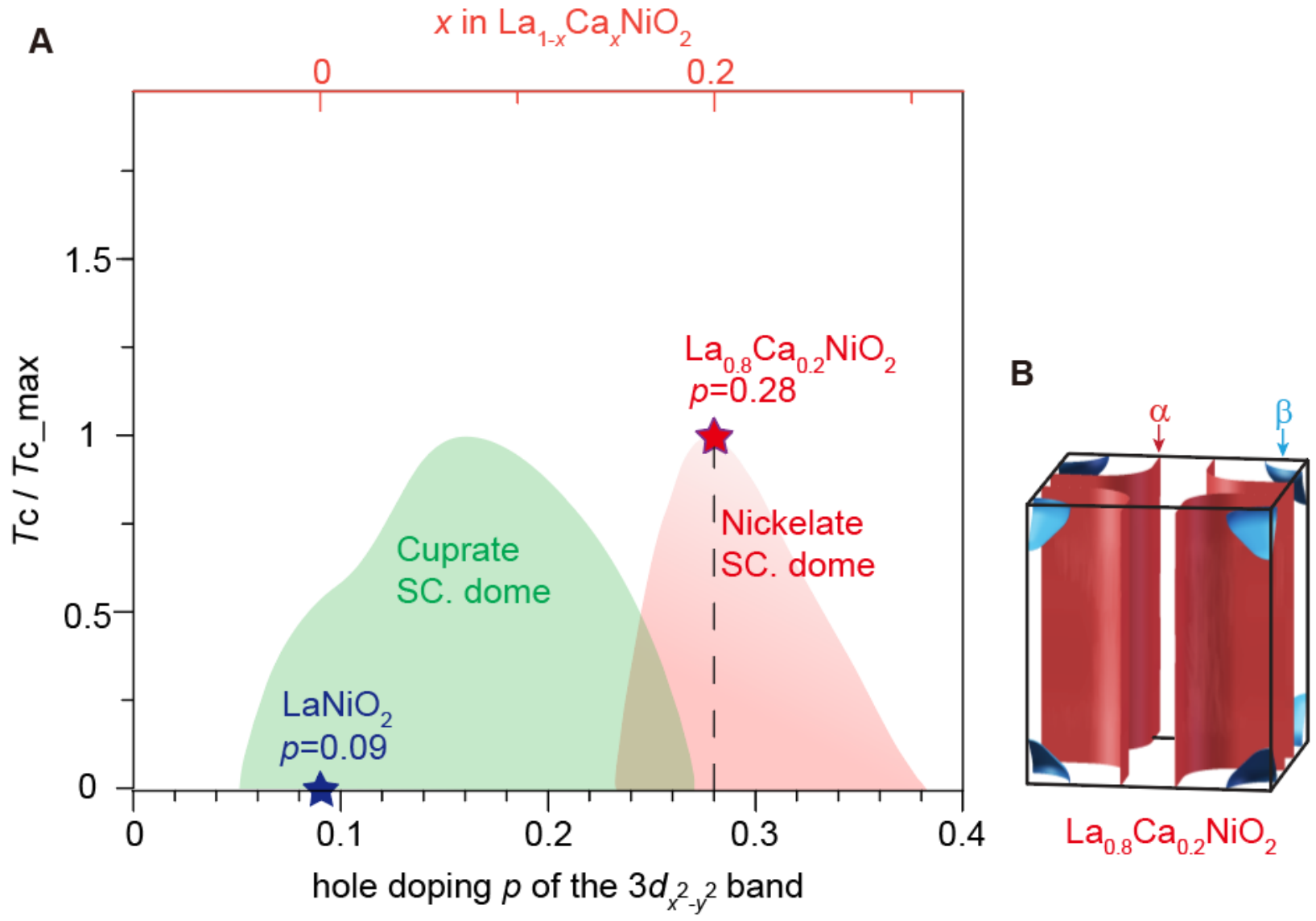

**Fig. 4. Substantial hole doping in the quasi-two-dimensional $\alpha$ band.** (**A**) Hole doping at the 3$d_{x2-y2}$ orbital plotted on the cuprate phase diagram. (B) Sketch of the measured three-dimensional Fermi surfaces observed in $La_{0.8}Ca_{0.2}NiO_2$.

## DISCUSSION

The quasi-2D $\alpha$ band dominated by Ni-3$d_{x2-y2}$ orbital and the 3D $\beta$ band around A qualitatively agree with the previous DFT results, where the orbital characters are predicted to be Ni-3$d_{x2-y2}$ for $\alpha$ band, and the mixing of La-5$d_{xy}$, Ni-3$d_{zx}/d_{yz}$, and interstitial-$s$ for $\beta$ band, respectively (Ref[34], Fig.S8). Nonetheless, some discrepancies with the calculated electronic structures are evident. Specifically, there is no electron pocket around Γ in $LaNiO_2$ (Fig.4B), and the $\beta$ electron Fermi pocket around the A point shows a slightly smaller size and a shallower band bottom than the calculations (Supplementary Fig.S7j). These observations consistently suggest that the self-doping effect is weaker than calculation prediction. The $\beta$ and $\alpha$ bands show different $E_F$ shifts upon doping (Fig.S7), indicating a non-rigid-band behavior. These discrepancies may be attributed to electron correlation effects overlooked in previous calculations, which potentially alters the dispersion [38]. Our ARPES data thus give a benchmark for further improving theoretical calculations.

In comparison to hole-doped cuprates, although the $\alpha$ band in IL nickelates generally captures the dispersion characteristics of the Zhang-Rice singlet band in cuprates, their superconducting doping ranges are far apart. As shown in Fig. 4A, the doping level of the $\alpha$ band in the parent compound $LaNiO_2$ already falls within the superconducting regime of cuprates, whereas the superconducting $La_{0.8}Ca_{0.2}NiO_2$ resides at the highly overdoped and non-superconducting regime of cuprates. The distinct superconducting hole-doping regimes of cuprates and IL nickelates may be rooted in the different nature of their electronic states. In cuprates, holes are predominantly doped into oxygen, and the states near the Fermi level primarily consist of oxygen ligand states. However, due to the large charge transfer gap in IL nickelates [21,39], the contribution of oxygen $p$ states is significantly less in IL nickelates compared to cuprates, and the $\alpha$ bands are predominantly dominated by the $d_{x^2-y^2}$ orbital (Supplementary Fig. S8). Furthermore, the $\beta$ band persists in the optimally doped superconducting $La_{0.8}Ca_{0.2}NiO_2$, suggesting the multiband nature of nickelate superconductors, distinct from that of cuprate superconductors.

To summarize, by reaching an unprecedented surface quality, we experimentally revealed the low-energy electronic structure of IL nickelates. Our ARPES measurements have revealed a large hole pocket $\alpha$ that bears resemblance to the Zhang-Rice singlet Fermi surface and dispersion in cuprates, and holes could be effectively introduced by Ca doping. The weak but finite self-doping effect, together with the highly hole-doped superconducting state, differs from the electronic structure of cuprates, posing constraints to theories. These findings clarified the fundamental issues on the electronic structure of IL nickelates, which paves the way toward the understanding of the superconductivity mechanism in IL nickelates. The observed large hole doping level in the superconducting $La_{0.8}Ca_{0.2}NiO_2$ is intriguing. It encourages the study of other IL nickelate systems, especially (Nd,Eu)$NiO_2$, which exhibits an additional doping level difference. The method developed for obtaining the single-crystalline surface of IL nickelates opens avenues for further surface-sensitive experimental studies on this family of compounds.

## MATERIALS AND METHODS

**Thin films growth.** Perovskite (La,Ca)$NiO_3$ thin films were grown on $TiO_2$-terminated $SrTiO_3$(001) substrates by oxide molecular beam epita0xy. A layer-by-layer growth mode is used, in which the A site element (La, Ca) and B site element (Ni) were deposited alternatively, while La and Ca were co-deposited to get uniform doping. The flux of each element was measured by quartz crystal microbalance (QCM), and then calibrated by Rutherford backscattering spectrometry (RBS) measurements. X-ray reflection (XRR) measurements were performed to further calibrate the absolute thickness of the films. XRD measurements were performed to optimize the growth conditions. After optimization, $LaNiO_3$ and Ca-doped $LaNiO_3$ were grown at 580°C under an ozone pressure of $5\times10^{-6}$mbar and $1.5\times10^{-5}$mbar, respectively. The 2D character of RHEED pattern is maintained during growth, indicating the single-crystalline and two-dimensional sample surface. The doping level calibrated by Rutherford backscattering spectrometry is then further checked by X-ray photoemission on the samples after ARPES studies.

***In-situ* reduction.** After growth, the precursor thin films were transferred *in-situ* to the pulsed laser deposition (PLD) chamber for reduction. Our PLD system is integrated with an atomic hydrogen gun, which generates atomic hydrogen by dissociating $H_2$ gas through plasma. IL $LaNiO_2$ and $La_{0.8}Ca_{0.2}NiO_2$ thin films were obtained by annealing perovskite precursors in an atomic hydrogen environment for 1~2 hours at 340°C, with a ramp rate of 15°C/min. During the reduction process, the $H_2$ gas flow rate was fixed at 3~sccm, and the chamber pressure was around $1.0\times10^{-5}$mbar. A metal shutter was used to prevent surface crystal structure degradation caused by exposure to $H^+$ (Fig.~1a). Under the optimized conditions, perovskite nickelates were transformed into IL nickelates, as confirmed by the X-ray diffraction pattern (see Fig.~1b and Fig S3a). Meanwhile, the fully-strained feature (Fig.~S3b) and the terraced surfaces were maintained in IL samples (Fig.~S4).

**ARPES measurements.** All the ARPES experiments were performed at the Shanghai Synchrotron Radiation Facility (SSRF). All samples were reduced *in-situ* and then transferred to beamline by vacuum suitcases and measured under an ultra-high vacuum better than $7\times10^{-11}$mbar. The SX-ARPES data and the complementary VUV-ARPES data were collected at beamline 09U and beamline 03U, respectively. In VUV ARPES experiments, we set the energy resolution power to 3000 for higher photon flux, which gives a typical energy resolution of 40~meV at 145eV photon energy. The estimated energy resolution of SX-ARPES is 100~meV at 250~eV, and 200~meV at 400~eV. The angle resolution is 0.1°.

More details on materials and methods can be found in Supplementary Materials[40-43].

## SUPPLEMENTARY DATA

Supplementary data are available at NSR online.

## ACKNOWLEDGEMENTS

We gratefully acknowledge the valuable discussion with Profs. Guangming Zhang, Jiangping Hu, and Hanghui Chen on theories, and Profs. Liang Qiao, Tong Zhang, Juan Jiang, and Ms. Yan Zhao on experiments. We thank Drs. Zhengtai Liu and Zhenhua Chen for the experimental support during the beamtime. We thank Profs. Liang Qiao, Yanwu Xie, Zhaoliang Liao, and Lingfei Wang, for providing samples grown by PLD in the early stage of this project. We thank Dr. Qingqin Ge for the XPS measurement.

## FUNDING

This work is supported in part by the National Natural Science Foundation of China (12074074, 12274085, 12174365), the National Key R&D Program of the MOST of China (2023YFA1406300), the New Cornerstone Science Foundation, the Innovation Program for Quantum Science and Technology (2021ZD0302803), and Shanghai Municipal Science and Technology Major Project (2019SHZDZX01), the China National Postdoctoral Program for Innovative Talents (BX20230078). Part of this research used Beamline 03U of the Shanghai Synchrotron Radiation Facility, which is supported by ME2 project (11227902) from National Natural Science Foundation of China.

## AUTHOR CONTRIBUTIONS

R.P., H.C.X., Y.F., Z.T.A., J.H.Y., S.L.T., and Z.H.C grew the perovskite films. X.D. and C.H.L. performed *in-situ* reduction and surface optimization. X.D., X.X.W., and C.H.L. set up the atomic hydrogen source. X.D., Z.T.A., J.H.Y., S.L.T., and S.S. performed various characterizations on the thin films. X.X.W., X.T.S., and N.G. in charge of the maintenance of the oxide MBE system. R.P., H.C.X., X.D., X.X.W., J.H.Y., M.Y.N.L., Y.F., C.H.L., and Z.H.C. performed ARPES measurements. R.P., H.C.X., and X.D. analyzed the ARPES data. Y.L.W. conducted the DFT calculations. D.L.F., R.P., and H.C.X. wrote the paper. D.L.F., R.P., and H.C.X. are responsible for the infrastructure, project direction, and planning.

**Conflict of interest statement.** None declared.

## REFERENCES

[1] Bednorz J G, Müller K A. Possible high *T*c superconductivity in the Ba-La-Cu-O system[J]. Eur Phys J B, 1986, 64(2): 189–193.

[2] Anisimov V, Bukhvalov D, Rice T. Electronic structure of possible nickelate analogs to the cuprates[J]. Phys. Rev. B, 1999, 59(12): 7901.

[3] Lee K-W, Pickett W. Infinite-layer $LaNiO_2$: $Ni^{1+}$ is not $Cu^{2+}$[J]. Phys. Rev. B, 2004, 70(16): 165109.

[4] Chaloupka J, Khaliullin G. Orbital order and possible superconductivity in $LaNiO_3/LaMO_3$ superlattices[J]. Phys. Rev. Lett., 2008, 100(1): 016404.

[5] Hansmann P, Yang X, Toschi A, et al. Turning a nickelate Fermi surface into a cupratelike one through heterostructuring[J]. Phys. Rev. Lett., 2009, 103(1): 016401.

[6] Han M-J, Wang X, Marianetti C, et al. Dynamical mean-field theory of nickelate superlattices[J]. Phys. Rev. Lett., 2011, 107(20): 206804.

[7] Li D, Lee K, Wang B Y, et al. Superconductivity in an infinite-layer nickelate[J]. Nature, 2019, 572(7771): 624–627.

[8] Osada M, Wang B Y, Goodge B H, et al. Nickelate superconductivity without rare-earth magnetism:$(La,Sr)NiO_2$[J]. Adv. Mater., 2021, 33(45): 2104083.

[9] Sun W, Li Y, Liu R, et al. Evidence for anisotropic superconductivity beyond pauli limit in infinite-layer lanthanum nickelates[J]. Adv. Mater., 2023, 35(32): 2303400.

[10] Zeng S, Li C, Chow L E, et al. Superconductivity in infinite-layer nickelate $La_{1-x}Ca_xNiO_2$ thin films[J]. Sci. Adv., 2022, 8(7): eabl9927.

[11] Osada M, Wang B Y, Goodge B H, et al. A superconducting praseodymium nickelate with infinite layer structure[J]. Nano Lett., 2020, 20(8): 5735–5740.

[12] Wei W, Vu D, Zhang Z, et al. Superconducting $Nd_{1-x}Eu_xNiO_2$ thin films using in situ synthesis[J]. Sci. Adv., 2023, 9(27): eadh3327.

[13] Lee K, Goodge B H, Li D, et al. Aspects of the synthesis of thin film superconducting infinite-layer nickelates[J]. APL Mater., 2020, 8(4).

[14] Gao Q, Zhao Y, Zhou X-J, et al. Preparation of superconducting thin films of infinite-layer nickelate $Nd_{0.8}Sr_{0.2}NiO_2$[J]. Chin. Phys. Lett., 2021, 38(7): 077401.

[15] Osada M, Fujiwara K, Nojima T, et al. Improvement of superconducting properties in $La_{1-x}Sr_xNiO_2$ thin films by tuning topochemical reduction temperature[J]. Phys. Rev. Mater., 2023, 7(5): L051801.

[16] Chow L E, Ariando A. Infinite-layer nickelate superconductors: A current experimental perspective of the crystal and electronic structures[J]. Front. Phys., 2022, 10: 834658.

[17] Parzyck C T, Anil V, Wu Y, et al. Synthesis of thin film infinite-layer nickelates by atomic hydrogen reduction: Clarifying the role of the capping layer[J]. APL Mater., 2024, 12(3): 031132.

[18] Kitatani M, Si L, Janson O, et al. Nickelate superconductors—a renaissance of the one-band Hubbard model[J]. npj Quantum Mater., 2020, 5(1): 59.

[19] Jiang M, Berciu M, Sawatzky G A. Critical nature of the Ni spin state in doped $NdNiO_2$[J]. Phys. Rev. Lett., 2020, 124(20): 207004.

[20] Zhang Y-H, Vishwanath A. Type-II $t$-$J$ model in superconducting nickelate $Nd_{1-x}Sr_xNiO_2$[J]. Phys. Rev. Res., 2020, 2(2): 023112.

[21] Hepting M, Li D, Jia C, et al. Electronic structure of the parent compound of superconducting infinite-layer nickelates[J]. Nat. Mater., 2020, 19(4): 381–385.

[22] Wu X, Di Sante D, Schwemmer T, et al. Robust $d_{x2-y2}$-wave superconductivity of infinite-layer nickelates[J]. Phys. Rev. B, 2020, 101(6): 060504.

[23] Nomura Y, Hirayama M, Tadano T, et al. Formation of a two-dimensional single-component correlated electron system and band engineering in the nickelate superconductor $NdNiO_2$[J]. Phys. Rev. B, 2019, 100(20): 205138.

[24] Gao J, Peng S, Wang Z, et al. Electronic structures and topological properties in nickelates $Ln_{n+1}$Ni_n$O_{2n+2}$[J]. Natl Sci. Rev., 2021, 8(8): nwaa218.

[25] Jiang P, Si L, Liao Z, et al. Electronic structure of rare-earth infinite-layer $RNiO_2$ ($R$=La, Nd)[J]. Phys. Rev. B, 2019, 100(20): 201106.

[26] Botana A S, Norman M R. Similarities and differences between $LaNiO_2$ and $CaCuO_2$ and implications for superconductivity[J]. Phys. Rev. X, 2020, 10(1): 011024.

[27] Lechermann F. Late transition metal oxides with infinite-layer structure: Nickelates versus cuprates[J]. Phys. Rev. B, 2020, 101(8): 081110.

[28] Karp J, Botana A S, Norman M R, et al. Many-body electronic structure of $NdNiO_2$ and $CaCuO_2$[J]. Phys. Rev. X, 2020, 10(2): 021061.

[29] Hu L-H, Wu C. Two-band model for magnetism and superconductivity in nickelates[J]. Phys. Rev. Res., 2019, 1(3): 032046.

[30] Werner P, Hoshino S. Nickelate superconductors: Multiorbital nature and spin freezing[J]. Phys. Rev. B, 2020, 101(4): 041104.

[31] Wang Z, Zhang G-M, Yang Y-f, et al. Distinct pairing symmetries of superconductivity in infinite-layer nickelates[J]. Phys. Rev. B, 2020, 102(22): 220501.

[32] Damascelli A, Hussain Z, Shen Z-X. Angle-resolved photoemission studies of the cuprate superconductors[J]. Rev. Mod. Phys., 2003, 75(2): 473.

[33] Xie B, Yang K, Shen D, et al. High-energy scale revival and giant kink in the dispersion of a cuprate superconductor[J]. Phys. Rev. Lett., 2007, 98(14): 147001.

[34] Sakakibara H, Usui H, Suzuki K, et al. Model construction and a possibility of cupratelike pairing in a new $d^9$ nickelate superconductor (Nd,Sr)$NiO_2$[J]. Phys. Rev. Lett., 2020, 125(7): 077003.

[35] Gu Y, Zhu S, Wang X, et al. A substantial hybridization between correlated Ni-d orbital and itinerant electrons in infinite-layer nickelates[J]. Commun. Phys., 2020, 3(1): 84.

[36] Been E, Lee W-S, Hwang H Y, et al. Electronic structure trends across the rare-earth series in superconducting infinite-layer nickelates[J]. Phys. Rev. X, 2021, 11(1): 011050.

[37] Keimer B, Kivelson S A, Norman M R, et al. From quantum matter to high-temperature superconductivity in copper oxides[J]. Nature, 2015, 518(7538): 179–186.

[38] Jiang R, Lang Z-J, Berlijn T, et al. Variation of carrier density in semimetals via short-range correlation: A case study with nickelate $NdNiO_2$[J]. Phys. Rev. B, 2023, 108(15): 155126.

[39] Goodge B H, Li D, Lee K, et al. Doping evolution of the Mott–Hubbard landscape in infinite-layer nickelates[J]. PProc. Natl Acad. Sci., 2021, 118(2): e2007683118.

[40] Blöchl P E. Projector augmented-wave method[J]. Phys. Rev. B, 1994, 50(24): 17953.

[41] Kresse G, Furthmüller J. Efficient iterative schemes for *ab* initio total-energy calculations using a plane-wave basis set[J]. Phys. Rev. B, 1996, 54(16): 11169.

[42] Perdew J P, Burke K, Ernzerhof M. Generalized gradient approximation made simple[J]. Phys. Rev. Lett., 1996, 77(18): 3865.

[43] Mostofi A A, Yates J R, Pizzi G, et al. An updated version of wannier90: A tool for obtaining maximally-localised Wannier functions[J]. Comput. Phys. Commun., 2014, 185(8): 2309–2310.